\documentclass[twocolumn,showpacs,prlprintnumbers,amsmath,amssymb]{revtex4}

\usepackage{graphicx}
\usepackage{dcolumn}
\usepackage{bm}

\begin{document}

\preprint{APS/123-QED}

\title{Anomalous coupling between topological defects and curvature}

\author{Vincenzo Vitelli}
\author{Ari M. Turner}
\affiliation{Department of Physics, Harvard University, Cambridge MA, 02138}%

\date{\today}

\begin{abstract}

We investigate a counterintuitive geometric interaction between
defects and curvature in thin layers of superfluids,
superconductors and liquid crystals deposited on curved surfaces.
Each defect feels a geometric potential whose functional form is
determined only by the shape of the surface, but whose sign and
strength depend on the transformation properties of the order
parameter. For superfluids and superconductors, the strength of
this interaction is proportional to the square of the charge and
causes all defects to be repelled (attracted) by regions of
positive (negative) Gaussian curvature. For liquid crystals in the
one elastic constant approximation, charges between $0$ and $4\pi$
are attracted by regions of positive curvature while all other
charges are repelled.

\end{abstract}

\pacs{Valid PACS appear here}

\maketitle The physics of topological defects on curved surfaces
plays an increasingly significant role in the engineering of
devices based on coated interfaces. \cite{Kami03,Nels02,baus03}.
Defects also affect the mechanical properties of some biological
systems, such as spherical viruses, whose shape is dependent on
the presence of disclinations in their protein shell \cite{Casp}.
Furthermore, the effects induced by a curved substrate on the
distribution of defects are not fully understood even in well
studied systems such as thin superfluid or superconducting films.
In this paper, we study simple continuum generalizations of the
plane XY model to frozen surfaces of varying curvature to gain a
broad understanding of the interaction between topological defects
and curvature.

The XY model is a simple setting in which particle-like objects
emerge from a more fundamental theory.  The basic degree of
freedom is an angle-valued function on the plane whose values vary
from $0$ to $2\pi$. These angles could represent the orientations
of interacting arrows. The interaction, which tends to align
neighboring arrows, results from the continuum free energy ${\cal
F}$ given by
\begin{equation}
{\cal F} = \frac{K}{2} \int d^{2}{\bf u} \!\!\!\!\! \quad
\left({\bf \nabla} \theta\left({\bf u} \!\!\!\!\! \quad \right)
\right)^{2} \!\!\!\! \quad , \label{eq:xy-plane}
\end{equation}
where the set of coordinates ${\bf u}=(x,y)$ label points on the
plane. Despite its simplicity, this model captures the main
properties of vortices in layers of superfluid $^{4}$He or thin
superconducting films when the field $\theta\left({\bf u}
\!\!\!\!\! \quad \right)$ is identified with the phase of the
collective wave function. In addition, the elastic energy of
Eq.(\ref{eq:xy-plane}) correctly describes liquid crystalline
phases for which the bond angle, $\theta\left({\bf u} \!\!\!\!\!
\quad \right)$, has periodicity $\frac{2\pi}{p}$ with $p \geq 3$.
For a solution of nematigens ($p=2$) and tilted molecules in a a
Langmuir film ($p=1$) two different elastic constants are
necessary to account for bend and splay deformations
\cite{deGennesbook}, but these are renormalized to the same value
at finite temperatures \cite{Nels-Pelc}. Besides its experimental
significance, the XY model is the cornerstone of our conceptual
understanding of topological defects, singular configurations of
the field $\theta\left({\bf u} \!\!\!\!\! \quad \right)$.

Like particles, defects have charges and a characteristic
Coulomb-like interaction. The charge $q$, a multiple of $\frac{2
\pi}{p}$, can be defined by the amount $\theta$ increases along a
path enclosing the defect's core. The force between two defects
located at positions ${\bf u}_{i}$ and ${\bf u}_{j}$ is determined
by the energy $E_{int}$ stored in the $\theta$ field:
\begin{equation}
E_{int} = K \!\!\!\! \quad q_{i} \!\!\!\!\! \quad q_{j} \!\!\!\!\!
\quad U({\bf u}_{i},{\bf u}_{j}) \!\!\!\! \quad , \label{eq:int-e}
\end{equation}
where the inter-particle potential $U({\bf u}_{i},{\bf u}_{j})$ is
proportional to the logarithm of the distance in the plane.

On a flat surface, thin layers of superfluids, superconductors and
liquid crystals can all be analyzed within the framework of
Eq.(\ref{eq:xy-plane}) \cite{Nels-Kost}. However, there is a
crucial difference between, say, the phase of the superfluid order
parameter and the angle that describes the local orientation of
liquid crystal molecules. The former transforms like a scalar
since the quantum mechanical phase does not change when the system
is rotated, while the latter represents a vector aligned to the
local direction of the molecules. Thus, a common boundary
condition for a liquid crystal is for the director to be tangent
to the boundary of the substrate. By contrast, no such constraint
exists for a $^4$He film because its wave function is defined in a
different space from the one in which the superfluid is confined.
This distinction is crucial on a curved surface. In the ground
state of a $^4$He film, the phase $\theta({\bf u})$ is constant
throughout the surface and the corresponding energy vanishes. The
free energy ${\cal F}_{s}$ to be minimized is a scalar
generalization of Eq.(\ref{eq:xy-plane}):
\begin{equation} {\cal F}_{s} = \frac{K}{2}\int
d^{2}u \!\!\!\!\!\! \quad \sqrt{g} \!\!\!\!\! \quad
g^{\alpha\beta}
\partial_{\alpha}\theta({\bf u}) \!\!\!\!\! \quad \partial_{\beta}\theta({\bf u}) \!\!\!\! \quad .
\label{eq:supfl-I}
\end{equation}
Here the set of coordinates ${\bf u}=(u_1,u_2)$ label points on
the surface while $\sqrt{g}$ is the determinant of the metric
tensor $g_{\alpha\beta}$. On the other hand, a constant bond angle
$\theta({\bf u})$ is not the ground state of the liquid crystal
because it is measured with respect to an arbitrary basis vector
${\bf E}_{\alpha}({\bf u})$ with $\alpha=1,2$. Indeed, it is not
possible to make the directions of the molecules parallel
everywhere on a curved space; the lowest energy state is attained
by optimally distributing the unavoidable bend and splay of the
vectors over the whole surface. The free energy functional ${\cal
F}_{v}$ to be minimized is a vector generalization of
Eq.(\ref{eq:xy-plane}) \cite{Davidreview}:
\begin{equation}
{\cal F}_{v} = \frac{K}{2}\int d^{2}u\sqrt{g} g^{\alpha\beta}
(\partial_{\alpha}\theta({\bf u}) - \Omega_{\alpha}({\bf
u}))(\partial_{\beta}\theta({\bf u}) - \Omega_{\beta}({\bf
u}))\!\!\!\! \quad , \label{eq:patic-ener}
\end{equation}
where $\Omega_{\alpha}({\bf u})$, the connection, compensates for
the rotation of the 2D basis vectors ${\bf E}_{\alpha}({\bf u})$
in the direction of $u_{\alpha}$.  Since the curl of
$\Omega_{\alpha}({\bf u})$ is equal to $G({\bf u})$ \cite{Nels87},
the integrand in Eq.(\ref{eq:patic-ener}) cannot be made to vanish
on a surface with non-zero Gaussian curvature
($\Omega_{\alpha}({\bf u})$ is a non-conservative field and hence
cannot be equal to $\partial_{\alpha}\theta$ everywhere). As the
substrate becomes more curved, the energy cost of this geometric
frustration can be lowered by generating defects in the ground
state even in the absence of topological constraints
\cite{Bowi-both,Vite03-II}.

In this letter, we introduce a novel coupling between a defect and
the varying curvature of the substrate which originates in a
conformal anomaly of the free energies of Equations
(\ref{eq:supfl-I}) and (\ref{eq:patic-ener}). This anomaly arises,
even at zero temperature, from imposing a constant cutoff, $a$,
localized at the core of each defect \footnote{We assume that $a$
is much smaller than the radius of curvature and hence independent
of the defect position.}. By contrast, finite temperature
conformal anomalies \cite{Poly81} are generated by the presence of
a short wavelength cutoff for the fluctuations in $\theta({\bf
u})$ at every point on the surface. A physical consequence of the
anomalous coupling discussed in this paper is that topological
defects in superfluids and superconductors interact with the
curvature in a radically different way from the case of liquid
crystal order \footnote{In both cases, the electrostatic
interaction between the defects $U({\bf u}_{i},{\bf u}_{j})$ is
given by the Green's function of the covariant Laplacian that is
not translationally invariant and depends on the shape of the
surface \cite{Vite03-II}.}.

\begin{figure}
\includegraphics[width=0.48\textwidth]{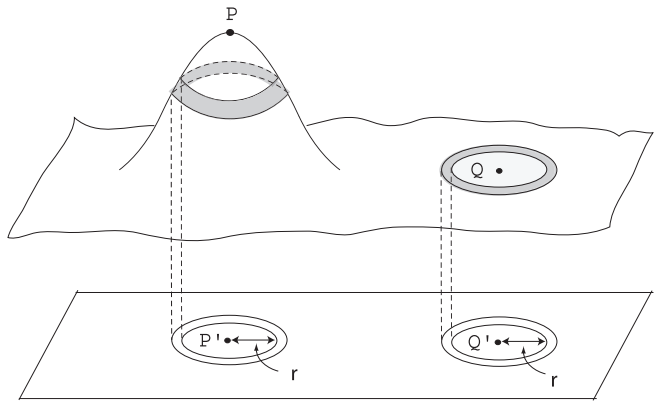}
\caption{\label{fig:intuition} A corrugated substrate and its
downward projection on a flat plane. The shaded strip surrounding
$P$ is more stretched than the one surrounding $Q$ despite their
projections onto the plane having the same area. The energy stored
in the field will be lower if the core of the defect is located at
$Q$ rather than $P$.}
\end{figure}

For thin layers of superfluids and superconductors, we prove that
the geometric interaction $E^{s}({\bf u}_{i})$ is given by:
\begin{equation}
E^{s}({\bf u}_{i})=- \!\!\!\!\!\! \quad \frac{K}{4\pi} \!\!\!\!
\quad q_{i}^{2} \!\!\!\!\! \quad V({\bf u}_{i}) \!\!\!\! \quad
,\label{eq:ecurv-s}
\end{equation}
where ${\bf u}_{i}$ and $q_{i}$ are respectively the position and
topological charge of the defect. The geometric potential $V({\bf
u})$ satisfies a covariant version of Poisson's equation where the
negative of the Gaussian curvature $G({\bf u})$ plays the role of
the charge density:
\begin{equation}
D_{\alpha}D^{\alpha}V({\bf u}) = G({\bf u}) \!\!\!\! \quad .
\label{eq:sol-chi}
\end{equation}
For an azimuthally symmetric surface such as the bump represented
in Fig. \ref{fig:intuition}, we can explicitly obtain $V({\bf
u})$, as a function of the radial distance from the top, by
employing a two dimensional analogue of Gauss' law
\cite{Vite03-II}. The resulting potential well $V(r)$ vanishes at
infinity and its width and depth are given respectively by the
linear size of the bump and its aspect ratio squared.
Eq.(\ref{eq:sol-chi}) has an elegant geometrical interpretation if
a set of coordinates is chosen so that the metric tensor is cast
in the form $g_{\alpha \beta}=\delta_{\alpha \beta}\exp
\left(-2\omega({\bf u})\right)$ \cite{Davidreview}. The conformal
factor $\omega \left({\bf u}\right)$ is controlled by the overall
shape of the surface and it satisfies the same Poisson
Eq.(\ref{eq:sol-chi}) as the geometric potential
\cite{Davidreview}. We therefore proceed with the identification
of $V({\bf u})$ with $\omega({\bf u})$ \footnote{This
identification is possible for corrugated planes flat at infinity.
The details of the analysis valid for deformed spheres will be
presented elsewhere \cite{Vite-Turn04}.}. This observation will be
the basis of our proof of Eq.(\ref{eq:ecurv-s}) which results in
the novel prediction that a vortex in a superfluid or
superconducting film is repelled (attracted) by positive
(negative) Gaussian curvature irrespective of its charge and sign.

For liquid crystals, the geometric interaction $E^{v}({\bf
u}_{i})$ contains an additional term discussed in previous
investigations of hexatic membranes \cite{Park96}, which arises
from the geometric frustration of the vector field. This term,
linear in $q$, happens to contain the same function $V({\bf u})$
as the conformal anomaly of Eq.(\ref{eq:ecurv-s}). When both
contributions are included, $E^{v}({\bf u}_{i})$ acquires an
unexpected dependence on the charge of the defect:
\begin{equation}
E^{v}({\bf u}_{i})= K \!\!\!\!\! \quad q_{i}
\left(1-\frac{q_{i}}{4 \pi} \right) \!\!\!\!\! \quad V({\bf
u}_{i}) \!\!\!\! \quad . \label{eq:ecurv-v}
\end{equation}
The interpretation of $V({\bf u})$ as a geometric potential and
the linear dependence on $q$ in the first term of
Eq.(\ref{eq:ecurv-v}) are consistent with the general belief that
a defect interacts logarithmically with the Gaussian curvature, as
an electrostatic particle would with a background charge
distribution. However, $E^{v}({\bf u}_{i})$ does not grow linearly
with the charge of the defect, as expected from the electrostatic
analogy. Instead, the geometric interaction peaks for a defect of
charge $2 \pi$ and eventually becomes repulsive for $q$ greater
than the critical charge $q_{c}=4\pi$. Furthermore, $E^{v}({\bf
u}_{i})$ is symmetric around $q=2\pi$ and does not simply flip
sign under charge inversion \footnote{Plots of field lines tangent
to the director for defects of different charge, $q$, (see Y.
Bouligand, in {\it Physics of Defects}, Les Houches 1980, pg. 683)
reveal a symmetry around $q=2 \pi$. The number of lobes in the
pattern of the field lines is the same for defects with $q=2
\pi(1+n/2)$ and $q=2 \pi(1-n/2)$, where $n$ is an arbitrary
integer.}.

The quadratic coupling has an intuitive explanation in the case of
azimuthally symmetric surfaces.
Consider a very thin superfluid film deposited on the surface
illustrated in Fig. \ref{fig:intuition} with a vortex of charge
$q$ placed on top of the bump. In order to calculate the energy
stored in the field, we only need to know that the superfluid
phase $\theta({\bf u})$ changes uniformly by $q$ along a
circumference of length $2\pi r$ centered on the defect.
Inspection of Eq.(\ref{eq:supfl-I}) reveals that the energy
density of the field in the shaded strip at distance $r$ is
proportional to $\left(\frac{q}{r}\right)^{2}$, where $r$ is the
distance to the singularity measured in the plane of projection
(see Fig. \ref{fig:intuition}). By vertically stretching the
surface, the amount of area in the shaded strip is increased with
respect to its projection on the plane, while the energy density
is unchanged. As a result, the total energy stored in the field is
greater when a vortex sits on top of a bumpy surface than when the
same vortex is located at the center of a flat disk of the same
area. Hence, it is energetically favorable for the vortex to
migrate to the flat portions of the surface. In this case, the
vortex is far away from the bump so that the total energy stored
in the field does not differ much from the flat plane result
\cite{Halperin-private}. For less symmetric surfaces, the
resulting geometric interaction will depend on the shape of the
entire surface as embedded in the metric tensor.

\begin{figure}
\includegraphics[width=0.48\textwidth]{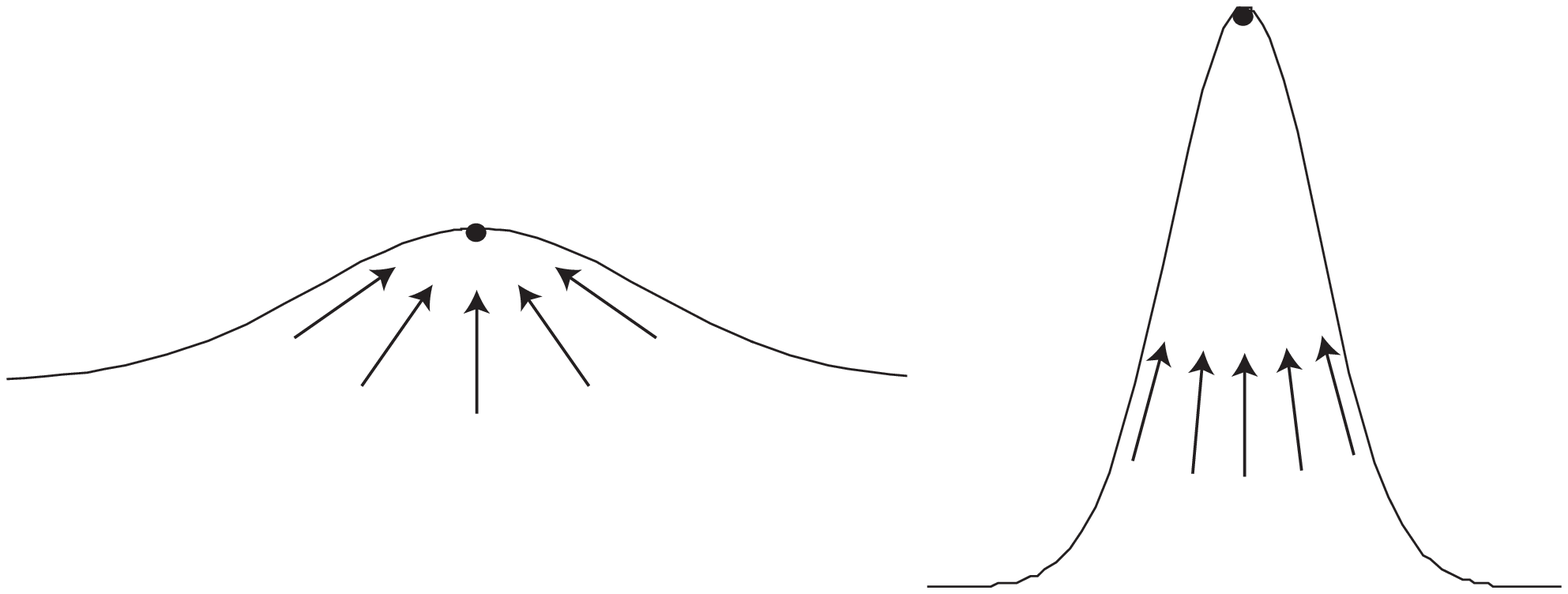}
\caption{Disclinations of charge $2\pi$ located on top of bumps
with different aspect ratios. The amount of splay in the liquid
crystal director on the taller bump is reduced and hence the
energy density is lower.} \label{fig-rice}
\end{figure}

The physical origin of the $linear$ coupling between defects and
curvature in Eq.(\ref{eq:ecurv-v}) is illustrated in Fig.
\ref{fig-rice} for a disclination of charge $2 \pi$ centered on a
bump. As the curvature of the bump is increased, the bend or splay
of the director of the liquid crystal decreases and hence the
energy stored in the vector field is reduced. As a result, this
linear coupling causes positively (negatively) charged defects to
be attracted by positive (negative) Gaussian curvature
\cite{Park96}. However, this mechanism competes with the repulsive
geometric interaction illustrated in Fig. \ref{fig:intuition} that
is at work also in the case of liquid crystal order. We note that
the linear coupling is absent for superfluids because, in
Eq.({\ref{eq:supfl-I}}), $\partial_{\alpha}\theta({\bf u})$ is not
coupled to a curvature dependent connection, $\Omega_{\alpha}({\bf
u})$, as it is in Eq.(\ref{eq:patic-ener}).

\begin{figure}
\includegraphics[width=0.48\textwidth]{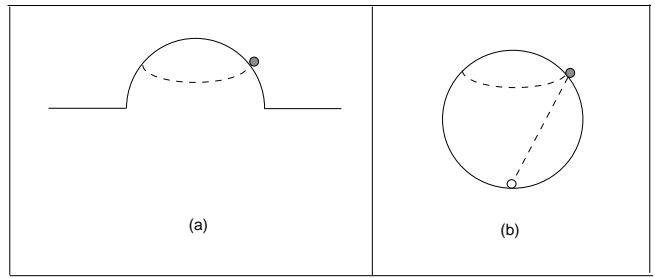}
\caption{(a) An isolated disclination on a deformed plane feels a
force that depends only on the enclosed Gaussian curvature. (b)
The deformed plane is compactified to the sphere by placing a
shadow defect at the south pole.} \label{fig-hemis}
\end{figure}

The critical value $q_{c}=4\pi$, where the single defect potential
$E^{v}({\bf u}_{i})$ changes sign, can be determined from simple
geometrical arguments. Consider an isolated disclination of charge
$q$ on a hemispherical cup placed on a flat plane. On account of
azimuthal symmetry, the force acting on the defect depends $only$
on the net Gaussian curvature enclosed by the circle on which it
is placed, see Fig. \ref{fig-hemis}a \cite{Vite03-II}. This
interaction is unchanged if we deform the outer region of the
plane and eventually compactify it to form a sphere as illustrated
in Fig. \ref{fig-hemis}b. In order to satisfy topological
constraints \footnote{The Poincare-Hopf theorem states that the
sum of the defects charges on a surface of genus $g$ is equal to
$4\pi(1-g)$ for a vector field and to $0$ for a scalar field
\cite{Davidreview}.}, we still need to place a shadow defect of
charge $(4\pi-q)$ at the south pole (the only position outside the
circle that does not destroy the azimuthal symmetry of the initial
problem). The curvature-defect interaction on the hemisphere is
thus reduced to the well known defect-defect interaction on the
sphere \cite{lube92}. The latter is proportional to $q(4\pi-q)$
and so is the curvature-defect interaction on the deformed plane
of Fig. \ref{fig-hemis}a, in agreement with Eq.(\ref{eq:ecurv-v}).
This provides evidence that a disclination of charge greater than
$4 \pi$ will be repelled from regions of positive curvature.

\begin{figure}
\includegraphics[width=0.45\textwidth]{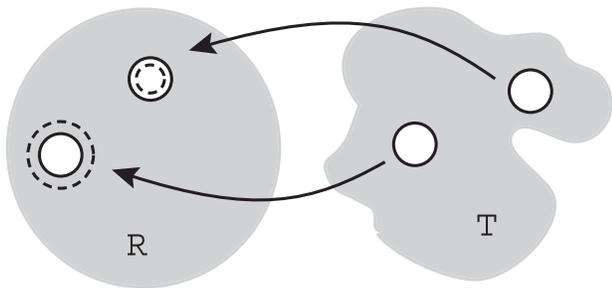}
\caption{Conformal mapping of the target surface $T$ onto the
reference space $R$. The continuous disks on both surfaces
represent the ``physical'' cores of constant radius $a$. The
dashed lines represent the position dependent images on $R$ of the
defect cores on $T$ with variable radii $a_i$. Note that the
energy stored in the annuli comprised by the dashed and continuous
lines in $R$ must be added or subtracted to $E_{R}$ to obtain
$E_{T}$.} \label{fig:core}
\end{figure}

We now present a derivation of the coupling between curvature and
defects in helium and superconducting films that employs the
method of conformal mapping, often adopted in electromagnetism and
fluid mechanics to simplify the boundary of complicated planar
regions. In this context, we use conformal mappings to relate the
complex task of finding the field energy on an arbitrarily
deformed target surface to an equivalent problem on a homogeneous
reference surface (see Fig. \ref{fig:core}). A conformal mapping
has two equivalent defining properties: angles map to equal
angles, and very small figures map to figures of nearly the same
shape. One can always find a conformal mapping from the target to
the reference surface \cite{Davidreview} such that $g_T=e^{- 2
\omega({\bf u})} g_R $, where $g_{T}$ and $g_{R}$ are the metric
tensors on the target and reference surfaces respectively. The
scaling factor $e^{\omega({\bf u})}$ varies with the position
${\bf u}$ on the target surface, so that larger figures are
inhomogeneously distorted when they are mapped from the target to
the reference surface. We choose the reference surfaces to be
undeformed and of the same topology as the target spaces (eg.,
$g_{R}=\delta_{\alpha \beta}$ for a corrugated plane). As a result
of the conformal transformation, defects on the target surface are
mapped onto a set of ``image defects'' on the reference surface.
The crucial property of the scalar free energy ${\cal F}_{s}$ is
its invariance under the rescaling of the metric by the conformal
factor \footnote{When $g_{T}$ is substituted in
Eq.({\ref{eq:supfl-I}}) the conformal factor $e^{- 2 \omega({\bf
u})}$ cancels out and $\sqrt{g_{T}} \!\!\!\!\quad
g^{\alpha\beta}_{T}=\sqrt{g_{R}} \!\!\!\!\quad
g^{\alpha\beta}_{R}$ \cite{Vite03-II}.}. However, the conformal
symmetry of ${\cal F}_{s}$ is broken upon introducing a short
distance cutoff $a$ that is necessary to prevent the energy from
diverging in the core of the defect. Because of the varying
scaling factor, the constant physical core radius $a$ is stretched
or contracted when projected on the reference space by an amount
dependent on the position of the defect (see Fig. \ref{fig:core}).
The radius of the image of the $i^{\rm th}$ core is $a_i=e^{
\omega({\bf u}_i)}a $. It is this conformal anomaly that is
responsible for generating the geometric interaction in
Eq.(\ref{eq:ecurv-s}). In fact, the energy of the defects in the
target space $E_{T}$ is equal to the energy of a configuration of
defects (on the reference surface) whose core radii are
position-dependent. This problem can be further transformed into
the simpler task of finding the energy $E_{R}$ for a set of
interacting defects with constant core radius $a$ plus an
effective geometric potential that accounts for the variation of
the core size with position. This geometric potential can be
derived with the aid of Fig. \ref{fig:core}.
If $a_i$ is smaller (larger) than $a$, the energies stored in the
annular regions indicated in Fig. \ref{fig:core} need to be added
(subtracted) from $E_{R}$ to obtain $E_{T}$. To calculate this
extra energy, we introduce a set of polar coordinates (r,$\phi$)
centered on the $i^{\rm th}$ defect. Near the defect of charge
$q_{i}$, the phase is given by $\theta \approx \frac{q_i}{2\pi}
\phi$ and the energy density is $\frac{K q_i^2}{8\pi^2 r^2}$. Upon
integrating it over the annulus comprised between $a$ and $a_i=e^{
\omega({\bf u}_i)}a$ (see Fig. \ref{fig-rice}), we obtain
\begin{equation}
E_{T}-E_{R}=-K \sum_{i=1}^{N_{d}}\frac{q_i^2}{4\pi} \!\!\!\!\quad
\omega({\bf u}_i)\!\!\!\!\quad , \label{eq:prsr}
\end{equation}
where $N_{d}$ is the number of defects. The energy $E_{R}$
accounts for defect-defect interactions since any potential felt
by a single defect would have to be constant because all points
are equivalent on the reference surface (undeformed sphere or
plane). Recalling that $\omega({\bf u})=V({\bf u})$, we recover
the result of Eq.(\ref{eq:ecurv-s}) with no dependence on the
microscopic physics because the core size $a$ drops out in
Eq.(\ref{eq:prsr}). In the case of liquid crystal order, the
contribution of the anomaly is simply added to the term linear in
$q$ as indicated in Eq.(\ref{eq:ecurv-v}) \cite{Vite03-II}.
Additional couplings result from abandoning the phase-only
approximation to the superfluid energy or from higher order terms
in a soft-spin Landau expansion. Generally, these energy
contributions vary inversely as the area of the surface and can be
made arbitrarily small by rescaling the surface, unlike the
geometric potential in Eq.({\ref{eq:prsr}}) \cite{Vite-Turn04}.

Experiments that test our predictions can be realized by coating a
bump with a thin layer of superfluid helium and rotating it around
its axis of symmetry so that a single vortex forms
\cite{Nelson-private}. The competition between the (repulsive)
geometric interaction and the confining parabolic potential
(generated by the rotation) would cause the equilibrium position
of the vortex to shift from the center of the bump if its height
exceeds a critical value. The vortex line could be detected by
trapping of electrons on its core \cite{Yarm82}. Other experiments
may detect an inhomogeneous distribution of thermally induced
defects resulting from the combined effect of the anomalous
coupling and the dependence of their Coulomb-like interaction on
the varying curvature.

We have demonstrated that the interaction between defects and
curvature in 2D XY-like models depends crucially on the nature of
the underlying order parameter and we have shown how to explicitly
derive the resulting geometric force from the shape of the
surface.

\textit{Acknowledgments:} We are grateful to D. R. Nelson and B.
I. Halperin for inspiring discussions and critical reading of the
manuscript and to R. E. Packard for conversations on the
experimental situation. We thank H. Chen, M. Forbes, J. Huang, Y.
Kafri and D. Podolsky for helpful comments. AMT has been supported
by a NSF graduate fellowship and VV by NSF under Grant No.
DMR-0231631 and by the Harvard Material Research Science and
Engineering Laboratory through Grant No. DMR-0213805.

\bibliography{anomalous}

\end{document}